\documentclass[twocolumn,aps,showpacs,pre,amsmath,amssymb,floatfix,longbibliography]{revtex4-2}

\usepackage{color}
\usepackage{multirow}
\usepackage{bm}
\usepackage{epsfig}
\usepackage{psfrag}
\usepackage[latin1]{inputenc}
\usepackage[english]{babel}
\usepackage{amsfonts}
\usepackage{amsmath}
\usepackage{upgreek}
\usepackage{enumitem}
\usepackage[dvipsnames]{xcolor}

\DeclareSymbolFont{bbold}{U}{bbold}{m}{n}
\DeclareSymbolFontAlphabet{\mathbbold}{bbold}

\begin{document}

\title{
	Random matrix analysis of deep neural network weight matrices
	}

\author{
	Matthias Thamm
	}  
\thanks{These authors contributed equally to this work.}
\author{
  Max Staats
	} 
\thanks{These authors contributed equally to this work.}
\author{
	Bernd Rosenow
	}  
\affiliation{
	Institut f\"{u}r Theoretische Physik, Universit\"{a}t
  Leipzig,  Br\"{u}derstrasse 16, 04103 Leipzig, Germany
	} 
	
\date{\today}
\begin{abstract}
		Neural networks have been used successfully in a variety of 
		fields, which has led to a great deal of interest in developing a 
		theoretical understanding of how they store the information needed to 
		perform a particular task.  
		We study the weight matrices of trained deep neural networks using methods 
		from random matrix theory (RMT)  and show that the  statistics of most of 
		the singular values follow universal RMT predictions.
		This suggests that they are random and do not contain system specific  
		information, which we investigate further by comparing the statistics of 
		eigenvector entries to the universal Porter-Thomas distribution. We  find 
		that for most eigenvectors the  hypothesis of randomness cannot be rejected, 
		and that only eigenvectors belonging to the largest singular values deviate 
		from the RMT prediction, indicating that they may encode learned information. In addition, a comparison with RMT predictions also allows to 
		distinguish networks trained in different learning regimes -- from lazy to rich 
		learning.
We analyze the spectral distribution of the large singular values 
		using the Hill estimator and find that the distribution cannot {in general} be characterized by a tail index, i.e.~is not of power law type.

\end{abstract}
 
\maketitle

\section{Introduction}

		The application of deep neural networks (DNNs)  
		to a wide range of problems 
		has  been tremendously successful  in recent years 	\cite{lecun2015, 
		Goodfellow.2016, Krizhevsky.2012, Silver.2017, Carleo.2019,
		bahri2020statistical}. Beyond the classification of images 
		\cite{Krizhevsky.2012}, DNNs 
		have been able to learn games beyond 
		human capabilities \cite{silver2016}, and have made significant progress in 
		solving hard problems like  protein folding \cite{jumper2021}. Given 
		these achievements, it is not surprising that neural networks are being
		applied to a variety of physics problems  as well \cite{Carleo.2019}. 
		Applications range from quantum state tomography \cite{Torlai.2018,
		Cao.2020}, locating classical \cite{Carrasquilla.2017} and quantum 
		phase transitions \cite{vanNieuwenburg.2017,Ch.2017,Broecker.2017,
		Huembeli.2018}, fluid turbulence \cite{lee1997application,
		jin2021nsfnets}, troposphere temperature prediction 
		\cite{chen2021physics} to classifying events in the Large Hadron Collider
		in search of physics beyond the standard model 
		\cite{duarte2018fast,guest2018deep}.

		Interestingly, successful neural networks are often highly 
		over-parametrized \cite{LeCun.1998,LeCun.1999,Pascanu.2014,Dauphin.2014,
		Goodfellow.2015,Neyshabur.2017,Soudry.2017,SiyuanMa.2018,Kawaguchi.2020,
		Krizhevsky.2009,Simonyan.2014,Zhai.2021}, such that one might have 
		doubts regarding their capability to generalize beyond the training 
		data set. Furthermore, DNNs 
		can effortlessly memorize
		large amounts of random training data \cite{lever2016points,Zhang.2021}, 
		but still  generalize well if there is a rule to be learned. According
		to the traditional concept of a ``bias-variance tradeoff''
		\cite{Geman.1992}, one would indeed expect these networks to overfit
		and fail in predicting unseen data. However, it turns out that their 
		generalization capability exhibits a so-called double descent behavior 
		\cite{Belkin.2019,Nakkiran19,Belkin.2020} as a function of the number
		of network parameters, such that they perform well in the highly 
		over-parameterized limit. The apparent contradiction between 
		over-parameterization and good generalization performance  is further 
		resolved by emerging evidence that ultra-wide neural networks are 
		inherently biased towards simple functions \cite{DePalma.2018,Valle.2018,
		bahri2020statistical,cohen2021learning}.
        For the analysis of such highly over-parameterized networks we use random matrix theory (RMT) \cite{Brody.1981,Guhr.1998,
		Mehta.2004,papenbrock2007colloquium,Weidenmuller.2009}  as a zero 
		information hypothesis, where deviations from RMT are indications of 
		system specific information. This approach has already proven to be useful for many other problems with inherent randomness such as the analysis of nuclei spectra \cite{wigner1951statistical,Guhr.1998,papenbrock2007colloquium,Weidenmuller.2009}, the investigation of stock market correlations \cite{Plerou.1999,Laloux.1999,Laloux.2000,Plerou.2002,Laloux.2000,
		Rosenow2002,schaefer2010,Bun2017}, and for the analysis of biological networks \cite{Luo2007,Deng2012}.

		Previously RMT has been applied
		to neural networks for estimating the asymptotic performance of single 
		layer networks \cite{Louart2018,Pennington2019}, and for analyzing the 
		generalization dynamics of linear networks \cite{lampinen2019analytic}. 
		Outliers and the random part of pre-activation covariance matrices were studied in \cite{Seroussi.2022}.
		Other work focused on the spectra of Jacobians at initialization 
		\cite{Pennington2018a} and on the eigenvalue distribution of the 
		Hessian of the loss matrix \cite{Baskerville.2021,granziol2020beyond}. 
		The spectral evolution of weight matrices during training was analyzed 
		in \cite{Martin.2021}, with the main results that for smaller or  older 
		networks  there is a scale in the singular value spectrum  separating signal 
		from noise, while in state of the art DNNs the spectral density exhibits a 
		power law tail, reminiscent of the spectrum of heavy-tailed random matrices 
		\cite{Cizeau1994,Tarquini2016}. These results were subsequently used 
		to assess the quality of pretrained DNNs \cite{Martin2021b} by computing 
		spectral  norms of weight matrices and by assessing the exponent of a power 
		law fit to the tail of the singular value spectrum.

		Here, we employ a variety of RMT tools to demonstrate that the weight matrices of  deep and over-parameterized neural networks are predominantly random. Specifically,  we compare the singular values of several DNNs  
		with the Wigner nearest neighbor spacing distribution and find that the 
		bulk of the singular values  follows the RMT prediction, even when the 
		networks are fully trained. This finding is also corroborated by the 
		analysis of the number variance of weight matrix singular value spectra.
		The advantage of analyzing these  RMT predictions for spectral correlations   is that they are universal properties of random matrices,  in  contrast to the distribution of singular values which depends sensitively on the  specific type of random matrix. 
		The idea that a large fraction of  singular values does not encode information is tested 
		by comparing the distribution of eigenvector components to the universal 
		Porter-Thomas distribution. Only for a small fraction of eigenvectors with large singular values do we find significant deviations 
		from the Porter-Thomas distribution, indicating that learned information is encoded in them. 
		
		In principle, it is also possible that information is stored in weights that match RMT predictions: For example, for a teacher-student setup with random teacher, one cannot locate information by an RMT analysis even for a perfect student of the same architecture -- thus, the absence of deviations from RMT predictions is no guarantee that no information was learned. This example shows that the success of the zero-information hypothesis approach depends on the structure of the data set on which the DNN is trained.  We show here, however, that the RMT analysis is particularly fruitful for image classification problems where networks trained on datasets such as MNIST, CIFAR, or ImageNet learn a rule in the images that can be encoded in a low-rank matrix such that the trained weights consist of a random bulk from initialization and a low-rank contribution.
		
		In addition, we train networks in different learning regimes from lazy networks for which weights barely change during training to rich networks where the final weights significantly differ from the initial ones \cite{Jacot.2018,Chizat.2019,Yang.2020,Fort.2020}.  We find that 
networks trained in the lazy regime follow RMT predictions, different from networks trained in the rich and in the intermediate regime. Thus, 		the form of the weight spectrum and the comparison of singular vector entries to the Porter-Thomas distribution allows to distinguish between the learning regimes where the best generalization performance is found in between both extremes.

		Furthermore, we apply the Hill estimator to study the question of whether large singular values of weight matrices can be described by a power law tail of the distribution  \cite{Martin.2021,Martin2021b}. 
		In contrast to previous results based on fitting the probability density function to a power law, both  the Hill estimator analysis and $p$-values for the significance of power law fits reveal  that the singular value distribution {for most of the considered cases} cannot be characterized by a tail exponent, implying that the tail is {generally} not described by a power law.

\section{Main results} \label{Sec:SummaryOfResults}
	 
	In this section, we illustrate our main results using 
	a 
	feedforward DNN  
	with three hidden layers, each containing 512 neurons {denoted as MLP512}. In the subsequent 
	sections, we provide more details and show that our results are valid more 
	generally for a variety of network architectures.  
        {The code needed to reproduce all results is open source and has been made available online \cite{CODE}.}
        
	We train {MLP512} networks on the CIFAR-10 dataset 
	\cite{Krizhevsky.2009}, which consists of images $\bm{x}^{(k)}$ (3072
	pixels each) and corresponding labels $\bm{y}^{(k)}$, which categorize the 
	images into 10 different classes.
	Therefore, in total the network has five layers with sizes 
	$\bm{n} = [3072,512,512,512,10]$.

		Except for the input layer, each layer $i$ has an associated weight 
		matrix ${\bf \sf W}_i$, a bias vector $\bm{b}_i$, and an activation 
		function $f_i(\bm{x})$ which we choose to be a rectified linear unit 
		(ReLU) $f(x)=\max(x,0)$ for the hidden layers and softmax 
		$f_{\rm out}(\bm{x})= \exp{\bm{x}}/\sum_i \exp{x_i}$ for the output 
		layer. The activation of the input layer $\bm{a}_0$ is defined as the 
		input image presented to the network
		\begin{align}
					\bm{a}_0^{(k)} &= \bm{x}^{(k)} \ , \\
					\intertext{and then the network propagates the
										activations through each layer such that the activation 
										in layer $i$ is given by}
					\bm{a}_i^{(k)} &= f_i( {\bf \sf W}_i\bm{a}_{i-1}^{(k)} + \bm{b}_i)\ .
 		\end{align}
	
		 Here, the $\bf \sf W_i$ are
		$n_i \times m_i$ weight matrices, with $n_i$ denoting the width of layer $i$ 
		and $m_i$ denoting the width of layer $i-1$. The output of the network is 
		defined as the activation of the last layer, using   the largest entry of 
		$\bm{a}_{\rm out}^{(k)}$ as prediction for the class of the input image 
		$\bm{x}^{(k)}$.  
		
		Before training, weights are initialized using the distribution Glorot 
		uniform \cite{Glorot.2010} while the biases are set to zero. Hence, the
		initial weight matrices are random matrices and obey RMT predictions (we note that 
		Glorot initialization and random Gaussian initialization do not differ in the singular value distribution and other statistical properties). 		
		The network's weights and biases are trained by minimizing the 
		cross-entropy cost function
		\begin{align}
					l(\bm{W},\bm{b}) = -\frac{1}{N}\sum_{k=1}^N \bm{y}^{(k)}  
					\cdot\ln(\bm{a}_{\rm out}^{(k)})
		\end{align}
	  on the training dataset using gradient descent on mini-batches of size 
		32. Further details of the training procedure are described in Appendix
		A. 

		\begin{figure}[t]
			\centering
			\includegraphics[width=8.6cm]{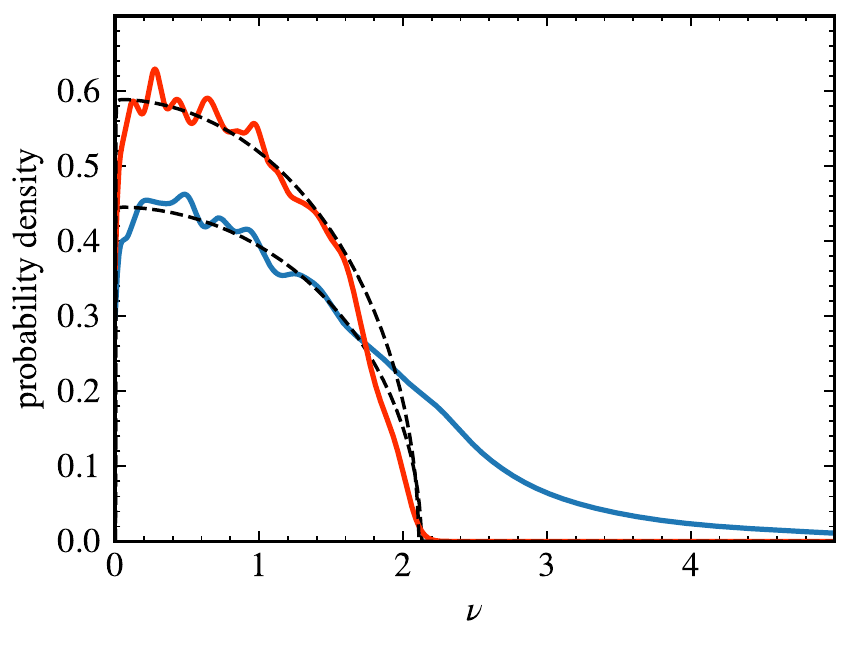}
			\caption{\label{Fig:Spectra}%
								Distribution of the singular values $\nu$ of the weight 
								matrix of the second hidden layer of the three hidden
								layer network {MLP512}. The spectral distributions are calculated by 
								broadening with a window size of $15$ singular values. The red 
								curve shows the distribution directly after random Glorot 
								initialization, and the blue line depicts the result after 
								fully training the network.
								The dashed, black lines are fits to the Marchenko-Pastur 
								distribution Eq.~\eqref{Eq:MarPas}.  
								After random initialization, the spectrum agrees well with the 
								RMT prediction, and even after training  the bulk of the 
								singular values still follows a {modified} Marchenko-Pastur distribution 
								with similar parameters.}
		\end{figure}
		
	By analyzing the singular values of weight matrices in trained DNNs  
	using methods  of RMT, 
	in the following 
	we argue that a large fraction of the weights remains random even after 
	training, while the learned information is encoded in relatively  few large 
	singular values and corresponding vectors. We focus on the second fully connected layer of {MLP512}, and obtain the singular values $\nu$ via 
	the singular value decomposition 
	${\bf \sf W} ={\bf \sf U} \, {\rm diag}(\nu) \, {\bf \sf V}^T$ of the 
	corresponding weight matrix $\bf \sf W$. Here, $\bf \sf U$ and $\bf \sf V$
	are orthogonal matrices, and ${\rm diag}(\nu)$ is a diagonal matrix 
	containing the singular values which we assume to be rank-ordered in the 
	following. 
	
	{We compare the trained weight matrices to their initialized states obtained by drawing entries of an $n \times m$  matrix independently and identically from a Glorot uniform distribution \cite{Glorot.2010} with variance $\sigma^2$.}
	The  singular values follow the Marchenko-Pastur distribution
	\cite{Marcenko.1967,Sengupta.1999,denby1991,denby1991}, 
	\begin{align}
		P(\nu) &= \begin{cases}
				\frac{n/m}{ \pi\tilde{\sigma}^2\nu}  
				\,\sqrt{(\nu_{\rm max}^2-\nu^2)(\nu^2-\nu_{\rm min}^2)} 
					 &  \nu \in [\nu_{\rm min},\nu_{\rm max}]\\
				 0 &  \text{else}
				\end{cases}   \label{Eq:MarPas}
	\end{align}
	where 
	$\nu_{\substack{\rm max\\ \rm min}} = \tilde{\sigma} (1\pm \sqrt{m/n})$ 
	and $\tilde{\sigma}=\sigma \sqrt{n}$. We assume without loss of 
	generality that $m \leq n$. 
	In the case of the second hidden layer weights, we have $n=m=512$. 
	While the distribution Eq.~\eqref{Eq:MarPas} is a parameter free 	
	description of the untrained network, the trained 
	network deviates from the Marchenko-Pastur law (see also the discussion in 
	\cite{Martin.2021}). 
	In the absence of a microscopic theory for the spectrum of a trained weight 
	matrix, we estimate its random part by  fitting  the empirical  spectra with a modified Marchenko-Pastur law in the 
	following way: we set $\nu_{\rm min}$ equal to the smallest empirical 
	eigenvalue, and then use $\nu_{\rm max}$ and $\sigma^2$ as fit parameters. 	This method has an additional free parameter as compared to the strict Marchenko-Pastur distribution, where the additional parameter can be understood as an estimate of the percentage of the spectrum that still follows the Marchenko-Pastur distribution.	
	\begin{figure}[t]
		\centering
		\includegraphics[width=8.6cm]{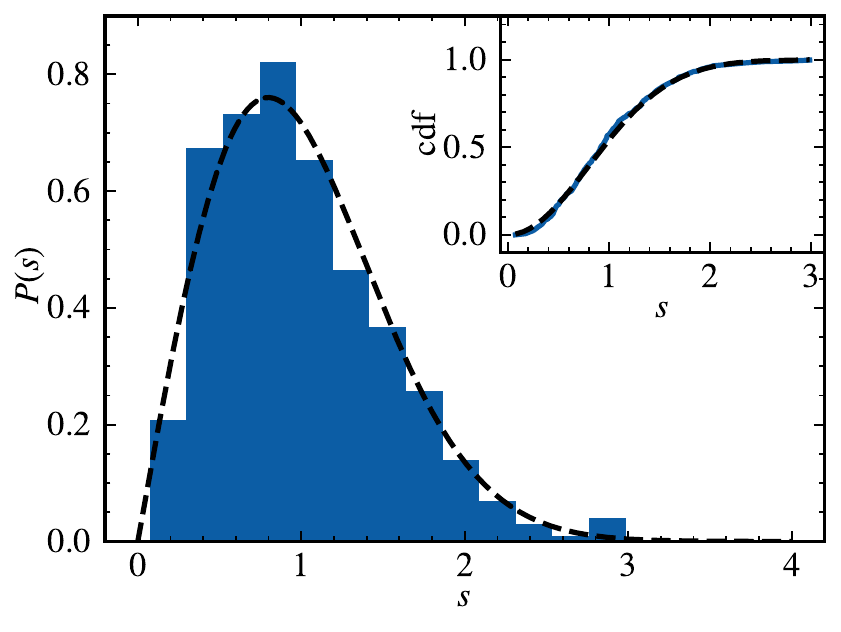}
		\caption{\label{Fig:SpacingsIntro}Spacing distributions of unfolded 
							singular values of the weight matrix of the second hidden 
							layer of {MLP512}. The main 
							panel depicts the probability density histogram and the 
							inset shows the cumulative distribution. In addition,  
							the RMT prediction (Wigner surmise)  Eq.~\eqref{Eq:wigner} 
							for matrices from the GOE is indicated by a dashed black line.  
							The prediction is confirmed both visually and by a 
							Kolmogorov-Smirnov test, which at the 40\% confidence level 
							cannot reject the hypothesis that the Wigner surmise is the 
							correct distribution.}
	\end{figure}

	Comparing the singular value spectra of the random weights at 
	initialization (red line in Fig.~\ref{Fig:Spectra}) with those of the 
	trained network (blue line in Fig.~\ref{Fig:Spectra}), it becomes 
	apparent that the bulk of the singular values still follows a {modified}
	Marchenko-Pastur distribution with similar parameters (dashed, black line
	in Fig.~\ref{Fig:Spectra}). In addition, there are some larger singular 
	values, which do not occur in the spectrum of the random control. 
	
	To further see that the majority of singular values of trained networks 
	are indeed random and do not encode information, we consider the 
	distribution of nearest-neighbor spacings of unfolded singular values. 
	Here, unfolding refers to	normalizing the mean density of states of the 
	singular values $\nu_i$ to unity, yielding the unfolded spectrum $\xi_i$ 
	\cite{Brody.1981,Guhr.1998,Mehta.2004,papenbrock2007colloquium,
	Weidenmuller.2009}. In contrast to the singular value distribution, which is 
	non-universal and depends on the system at hand, the spacing distribution is a 
	universal property of random matrices. For real random matrices in the 
	universality class of the Gaussian orthogonal ensemble (GOE), the level	
	spacings $s_k = \xi_{k+1}-\xi_k$, i.e.~the differences between neighboring 
	singular values,   are distributed according to the Wigner surmise 
	\cite{Brody.1981,Guhr.1998,papenbrock2007colloquium,wigner1951statistical,
	Mehta.2004,Weidenmuller.2009}
	\begin{align}
		 P_{\rm GOE}(s) = \frac{\pi s}{2}\,\exp\left(-\frac{\pi}{4}\,s^2\right)
				\ . \label{Eq:wigner}
	\end{align}
	The nearest neighbor spacings of the weight matrix singular values are in 
	excellent agreement with the RMT prediction Eq.~\eqref{Eq:wigner} even 
	after training the networks (Fig.~\ref{Fig:SpacingsIntro}). This is 
	supported by a Kolmogorov-Smirnov test of the empirical data against 
	Eq.~\eqref{Eq:wigner}	that cannot reject the null hypothesis even at a 
	significance level as high as $\alpha=0.40$.

	\begin{figure}[t]
		\centering
		\includegraphics[width=8.6cm]{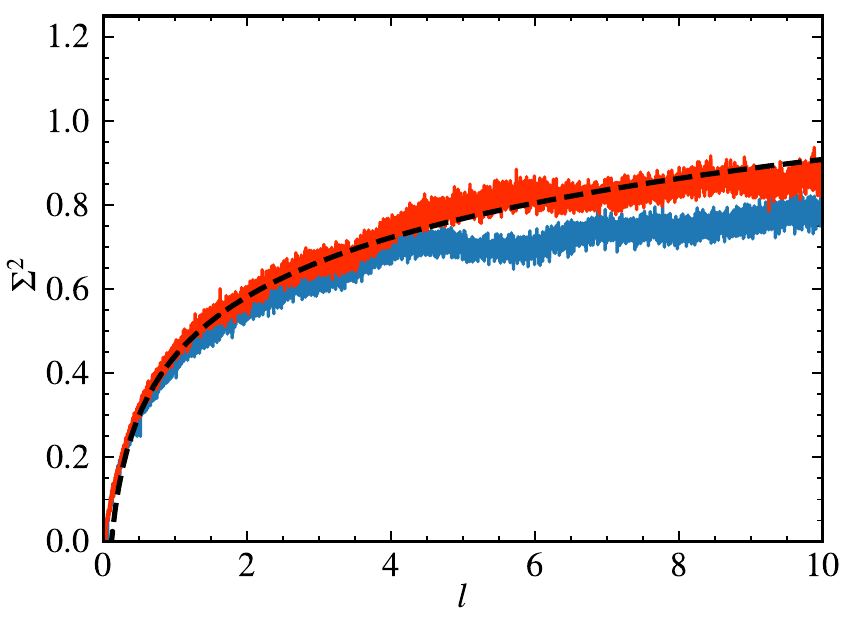}
		\caption{\label{Fig:Rigidity}Level number variance $\Sigma^2$ of singular
						  values of the weights for the second hidden layer of {MLP512}, after initialization (red) and after 
							fully	training the network (blue). The dashed black line 
							is the theory prediction for the GOE.
							We find that even after training, the level number variance  
							grows only logarithmically, as predicted by RMT.}
	\end{figure} 	 
	
	Another prediction of RMT that allows to test the random nature of weight 
	matrices is the level number variance, which is sensitive to long range 
	correlations in the spectrum. The number variance describes fluctuations 
	in the number of unfolded singular values $N_{\xi_i}(l)$ in intervals of 
	length $l$ around each singular value $\xi_i$  
	\begin{align}
		\Sigma^2(l) = \langle (N_{\xi}(l)-l)^2 \rangle_\xi\ . \label{Eq:LNV}
	\end{align}
	For random matrices from the GOE universality class, the level number 
	variance depends on the interval width $\ell$ according to 
	$\Sigma^2(l) \propto  \ln(2\pi l)$ in the regime $l \gtrsim 1$ \cite{Guhr.1998,Mehta.2004,
	papenbrock2007colloquium,Weidenmuller.2009} (dashed, black line in 
	Fig.~\ref{Fig:Rigidity}). This is in very good agreement with empirical  results for 
	the trained example network (Fig.~\ref{Fig:Rigidity}). In particular, there 
	are only small differences between randomly initialized weights (red lines in 
	Fig.~\ref{Fig:Rigidity}) and the fully trained weight matrices.

	\begin{figure}[t]
		\centering
		\includegraphics[width=8.6cm]{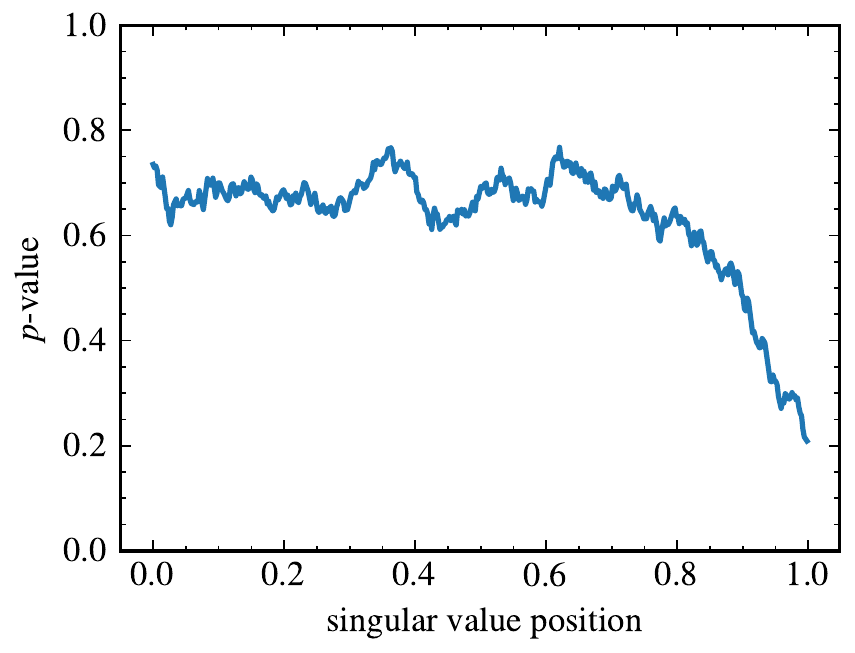}
		\caption{\label{Fig:MLP_eigenvectorPVals}%
							Randomness of eigenvectors as a function of the singular value 
							position in the spectrum: we quantify the agreement with the RMT 
							Porter-Thomas distribution by computing the $p$-value of a 
							Kolmogorov-Smirnov test  on the entries of the eigenvectors of 
							${\bf \sf W}^\dagger {\bf \sf W}$	for the second hidden layer
							of MLP512.  
							On the $x$ axis, we plot the positions according to the rank 
							ordered singular values, such that 0 corresponds to the 
							smallest and 1 to the largest singular value of the weight
							matrix.
							The results are averaged over neighboring singular values
							with a window size of 15.
							For large  singular values, the hypothesis of random eigenvectors 
							is rejected, indicating that information is stored in these 
							singular values and eigenvectors. 
							}
							
	\end{figure} 

	In addition, we consider the normalized eigenvectors of the $m\times m$ matrix 
	${\bf \sf W}^\dagger {\bf \sf W}$ (the right singular vectors of 
	${\bf \sf W}$), whose components in the case of a random matrix are 
	described by the Porter-Thomas distribution 
	\cite{Guhr.1998,papenbrock2007colloquium,Weidenmuller.2009}, 
	i.e.~a Gaussian distribution {with  mean of zero and a standard deviation which is fixed by the normalization of the vector's length to unity} as
	$1/\sqrt{m}$ . To verify whether the 
	observation that most singular values of trained networks are random also 
	carries over to the associated eigenvectors, we test the empirical 
	distribution of the entries of each eigenvector against such a Gaussian
	distribution using a Kolmogorov-Smirnov test. 
	If the test returns a large 
	$p$-value, the hypothesis of a Gaussian distribution cannot be rejected, which we interpret as an indication that the corresponding vector contains only noise. The rejection of the Gaussian hypothesis for small $p$-values is an indication for stored information.  
	To reveal a trend in the data, we average the resulting $p$-values over 
	neighboring singular values with a window size of $15$.
	Indeed, we find that most eigenvectors are random, especially those 
	belonging to small singular values (Fig.~\ref{Fig:MLP_eigenvectorPVals}).  
	For large singular values, we observe a decrease of the $p$-values, 
	indicating that relevant information is stored in the corresponding 
	eigenvectors, which is also consistent with the results 
	\cite{Martin.2021}.

\begin{figure}[t]
		\centering
		\includegraphics[width=8.6cm]{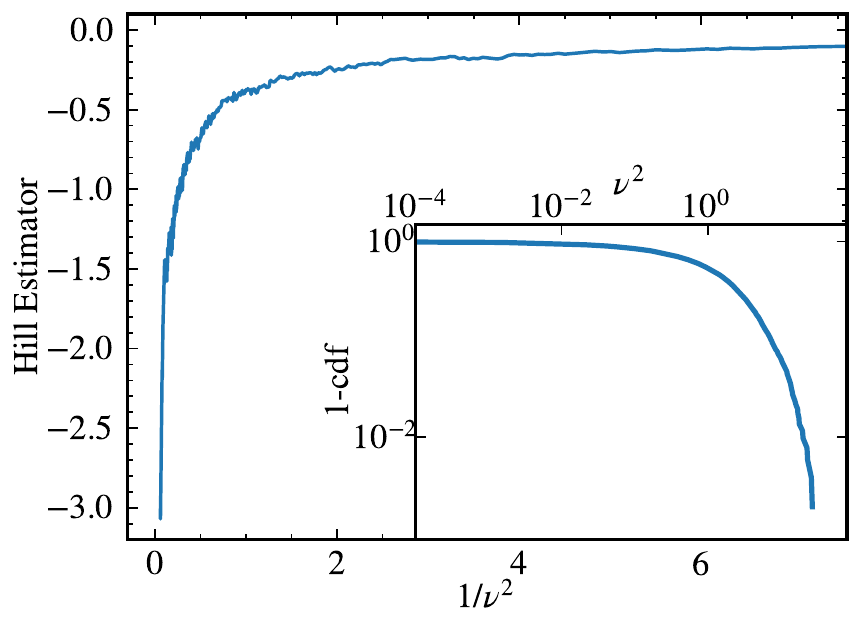}
		\caption{\label{Fig:Hill_512}
						Estimate  of the tail exponent of the singular value spectrum of  {the second hidden layer of MLP512} obtained by averaging the 
						inverse local slopes obtained via a Hill estimator  with window size $a=20$.
						 The absence of a flat plateau region shows that no power 
						law is present, even though this is not immediately evident in  
						a double logarithmic plot of the cumulative distribution (inset). 						
							}
							
	\end{figure}

We further analyze the tail region of the singular value distribution, which was recently described in terms of power law fits to the probability distribution function \cite{Martin2021b,Martin.2021}. To this end, we consider 
	the eigenvalues $\lambda=\nu^2$ of  the matrix ${\bf \sf W}^\dagger {\bf \sf W}$, 
	and analyze a potential power law decay of the cumulative distribution function using the Hill estimator \cite{hill1975} obtained by 
	averaging the inverse local slopes of the log-log cumulative distribution. 
	A power law tail would manifest itself in an extended flat region of a Hill plot, which is absent in  Fig.~\ref{Fig:Hill_512}.  
	In Sec.~\ref{Sec:Hill} we extend this analysis to  a variety of over-parameterized DNNs and do not find  evidence for a power law behavior in most of them, in contrast to the characterization of DNNs in terms of power law fits to the distribution function of weight matrix singular values \cite{Martin2021b,Martin.2021}. These findings are consistent with an analysis of $p$-values for power law fits.

	\begin{figure}[t!]
		\centering
		\includegraphics[width=8.6cm]{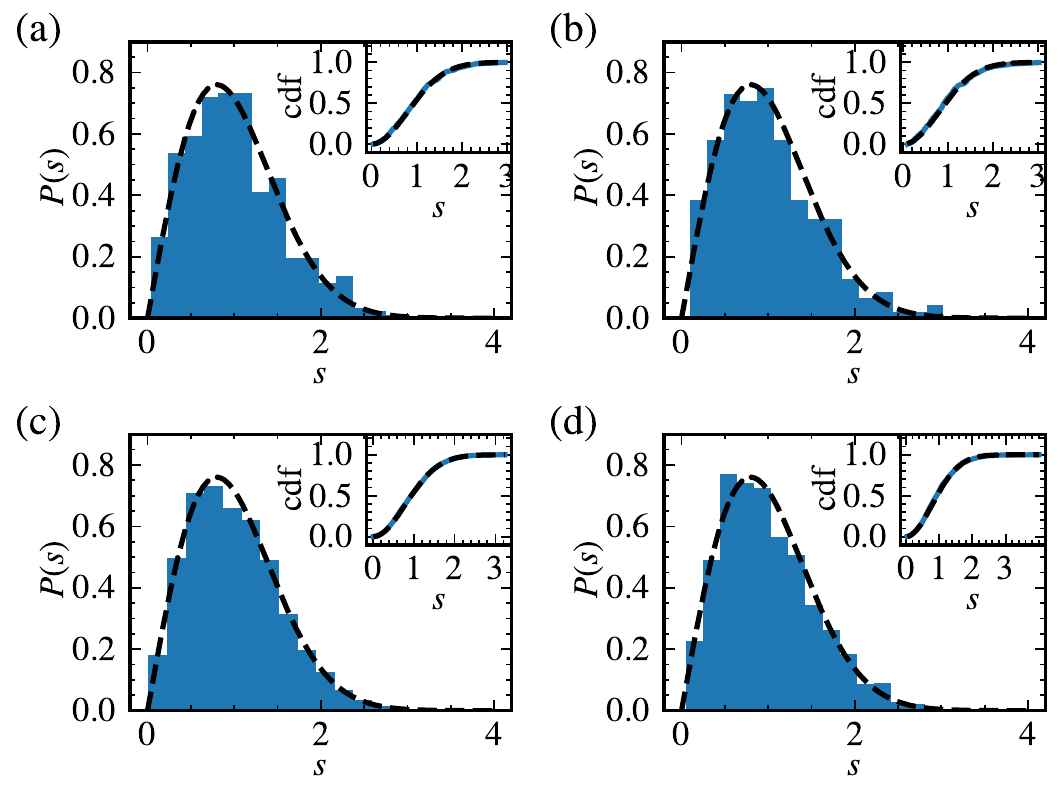}
		\caption{\label{Fig:Spacings} Nearest neighbor spacing distributions of 
							unfolded singular values of weight matrices for various neural 
							networks. The main panels depict the probability density 
							histograms and the insets show the cumulative distribution 
							functions. In addition, we depict the	Wigner surmise theory 
							prediction Eq.~\eqref{Eq:wigner} for the GOE with dashed, black 
							lines. (a) Results for the second hidden layer weight matrix of {MLP1024}. 
							In addition, results for (b) the second convolutional layer 
							in the CNN miniAlexNet, (c) the second fully connected layer 
							in AlexNet, and (d) for the third dense layer in VGG19.
							In all cases there is excellent visual agreement
							with the RMT predictions. This is further supported by 
							Kolmogorov-Smirnov tests which cannot reject the 
							null hypothesis at a significance level of 
							(a) 81\%, (b) 85\%, (c) 31\%, and (d) 96\%.  
							}
	\end{figure}

\section{Universality of level spacing distribution} \label{Sec:UniLS}

	\begin{table*}[t]
			\caption{\label{Tab:KSTests}Kolmogorov-Smirnov test results of the 
								distribution of unfolded singular value spacings of the 
								weight matrices against the Brody distribution with 
								$\beta=1$. Rejection of the null hypothesis is based on the
								$\alpha=0.05$ significance level. The p-value indicates how 
								likely it is to obtain a distribution with at least as much
								cumulative density function deviation as the one tested for 
								drawing random numbers from a Brody distribution with 
								$\beta=1$. In addition, we show the	results of a fit of a 
								Brody distribution with free parameter $\beta$ to the 
								cumulative density function of the computed level spacings. 
								The error was determined by a bootstrap sampling method.
								We find excellent agreement with the Wigner
								surmise for a variety of network architectures.}
			\begin{tabular}{|c|r|c|c|c|c|c|c|c|c|c|}
			\hline
			\multirow{2}{*}{network}&	 
			\multicolumn{3}{|c|}{reject null hypothesis?}&\multicolumn{3}{|c|}{ks-test p-value}&\multicolumn{3}{|c|}{Brody $\beta$ from fit} \\
															&						  
															layer 1 & layer 2 & layer 3 &  layer 1 & layer 2 & layer 3			&   layer 1 & layer 2 & layer 3	\\ 
			\hline
			\multirow{1}{6.2cm}{MLP512 (seed 1, Sec.~II) } 
															&			 	
															no&no&no						&  0.347  & 0.401   &	 0.812  &   $0.79\pm 0.09$ &	$1.01\pm 0.10$ &	$1.04\pm 0.12$\\\hline
			\multirow{1}{6.2cm}{MLP512 (seed 2) } 
															&			 	
															no&no&no						& 0.993   & 0.421   &	0.844    &   $0.99  \pm 0.11  $ &	$ 1.01 \pm  0.11 $ &	$ 0.99 \pm 0.11 $\\  \hline
			\multirow{1}{6.2cm}{MLP512 (seed 3) } 
															&			 	
															no&no&no						& 0.768   & 0.784   &	0.863    &   $ 0.95 \pm 0.11 $ &	$ 0.93 \pm 0.10  $ &	$ 0.92 \pm 0.09 $\\  
			\hline
			\multirow{1}{6.2cm}{MLP1024  } 
															&			 	
															no&no&no						&  0.799  & 0.812   &	 0.792  &   $0.94\pm0.07$ &	$0.91\pm0.11$ &	$1.03\pm0.10$\\  
			\hline
			\multirow{1}{6.2cm}{miniAlexNet (second conv.\@ layer)} 
															&		 	
															\multicolumn{3}{|c|}{no}	&  \multicolumn{3}{|c|}{0.859}  &  \multicolumn{3}{|c|}{$0.85 \pm 0.14$} \\ 
			\hline
			\multirow{1}{6.2cm}{AlexNet (dense layers) -- \texttt{torch}} 
			&			 
			no&no&no
			&   	0.670  &	0.229  &	0.160  
			&  $0.96\pm0.04$ &	$0.95\pm0.04$	& $0.83\pm0.07$  \\
			\hline
			\multirow{1}{6.2cm}{VGG16 (dense layers) -- \texttt{tensorflow}} 
															&			   
															no&no&no						&   	0.923  &	0.312 &	0.309   &  $  0.99\pm0.04$ &	$ 0.92\pm 0.04 $	& $ 0.92\pm0.07 $  \\
			\hline
			\multirow{1}{6.2cm}{VGG19 (dense layers) -- \texttt{torch}} 
															&			   
															no&no&no						&   	0.376  &	0.652 &	0.557   &  $  0.97\pm0.04$ &	$ 0.95\pm 0.04 $	& $ 0.92\pm0.07 $  \\
			\hline
			\end{tabular}
		\end{table*}
		To demonstrate that the results presented in the previous section are 
		typical	of trained DNNs, 
		we consider several networks with
		different architectures and sizes here and in the following sections: a) a fully 
		connected feedforward network with layers of size [3072, 1024, 512, 512, 10] {denoted as MLP1024}
		and b) a convolutional network called miniAlexNet consisting of two 
		convolutional layers followed by three dense layers, both trained on 
		the CIFAR-10 \cite{Krizhevsky.2009} dataset. In addition, we analyze 
		the two large networks c) AlexNet \cite{Krizhevsky.2017} and d) VGG19
		\cite{Simonyan.2014}, whose models trained on the ImageNet 
		\cite{Deng.2009} dataset are available via \texttt{pytorch} \cite{pytorch.2019} and \texttt{tensorflow} 
		\cite{Tensorflow.2015}. More details on the definition 
		of the networks, training parameters, and performance of the fully
		trained networks can be found in Appendix A.

		In order to compare the various  RMT predictions with the properties of 
		empirical weight matrices, we compute  their singular value decompositions. 
		While this is straightforward for dense layers, in the case of convolutional 
		layers we first reshape the four dimensional weight tensors to a	rectangular shape 
		(for details see App.~B) and then compute their singular values and vectors. 
		To obtain  smooth probability densities for the singular value spectra,  
		we perform a Gaussian broadening \cite{Brack.1972,Mehta.2004} by 
		approximating the probability density as a sum of Gaussian functions 
		centered around each of the $m$ singular values $\nu_k$ with 
		widths $\sigma_k = (\nu_{k+a}-\nu_{k-a})/2$, where $a$ is	the window 
		size of the broadening \cite{Bruus.1996,Plerou.2002}
	  \begin{align}
			P(\nu) \approx \frac{1}{m}\sum_{k=1}^m \frac{1}{\sqrt{2\pi\sigma_k^2}} \exp\left(
					-\frac{(\nu-\nu_k)^2}{2\sigma_k^2}\right) \label{Eq:GbroadPv}\ .
	  \end{align}
		To compare the RMT prediction Eq.~\eqref{Eq:wigner} with the level 
		spacing of the networks weights, we have to unfold the singular value 
		spectrum first.
		Here, unfolding is a transformation that maps the singular values 
		$\nu_i$ to	uniformly distributed singular values $\xi_i$ 
		\cite{Brody.1981,Guhr.1998,Plerou.2002,Mehta.2004,Weidenmuller.2009}. 
		For this purpose, we first determine the probability density $P(\nu)$ using 
		Eq.~\eqref{Eq:GbroadPv} and calculate the corresponding cumulative 
		distribution
		\begin{align}
			F(\nu) = m\int_{-\infty}^\nu P(x)\,dx\ . \label{Eq:cdf}
		\end{align}
		The unfolded singular values are  defined as $\xi_i = F(\nu_i)$. We then 
		obtain the spacings of the unfolded and sorted singular values $\xi_i$ 
		via
		\begin{align}
				s_k = \xi_{k+1}-\xi_k \ .
		\end{align}
		We find excellent agreement with the RMT predictions for all network
		architectures and layers (Fig.~\ref{Fig:Spacings}). This is also 
		supported by Kolmogorov-Smirnov tests (see Tab.~\ref{Tab:KSTests}) with 
		null hypothesis that the distribution is described by the Wigner 
		surmise Eq.~\eqref{Eq:wigner}. 
		We infer from the tests that for all network types, for fully connected 
		as well as for convolutional layer, the null hypothesis cannot be 
		rejected on the $\alpha=0.05$ significance level. 
		The $p$-values of the Kolmogorov-Smirnov 
		tests (Tab.~\ref{Tab:KSTests}) even show that in many cases a rejection
		for much higher $\alpha$ is also not possible.

		Furthermore, we consider the more general case of a Brody distribution
		\cite{Brody.1981,Guhr.1998,Weidenmuller.2009}
		\begin{align}
					P_{\rm Br}(s) = B(1+\beta)s^\beta \exp(-B s^{1+\beta})\ , 
					\label{Eq:Brody}
		\end{align}
		with $B=\{\Gamma([\beta+2]/[\beta+1])\}^{1+\beta}$. For $\beta=1$ this
		reduces to the Wigner surmise Eq.~\eqref{Eq:wigner}. Fits of the Brody
		distribution Eq.~\eqref{Eq:Brody} to the data show very good agreement 
		with $\beta\sim 1$ (see Tab.~\ref{Tab:KSTests}). To obtain $\beta$ and 
		the error estimate, we use bootstrap sampling \cite{Efron.1979,
		Efron.1998,Efron.2003}, fit the Brody distribution to each sample, and
		then calculate $\beta$ as the mean and the error as the standard 
		deviation of all fit results.

		The level spacing results suggest that the majority of weights is random
		even after training. This indicates that the weights, even in the case of
		the large networks trained on ImageNet, have rather low information	
		density.

	\begin{figure}[t!]
		\centering
		\includegraphics[width=8.6cm]{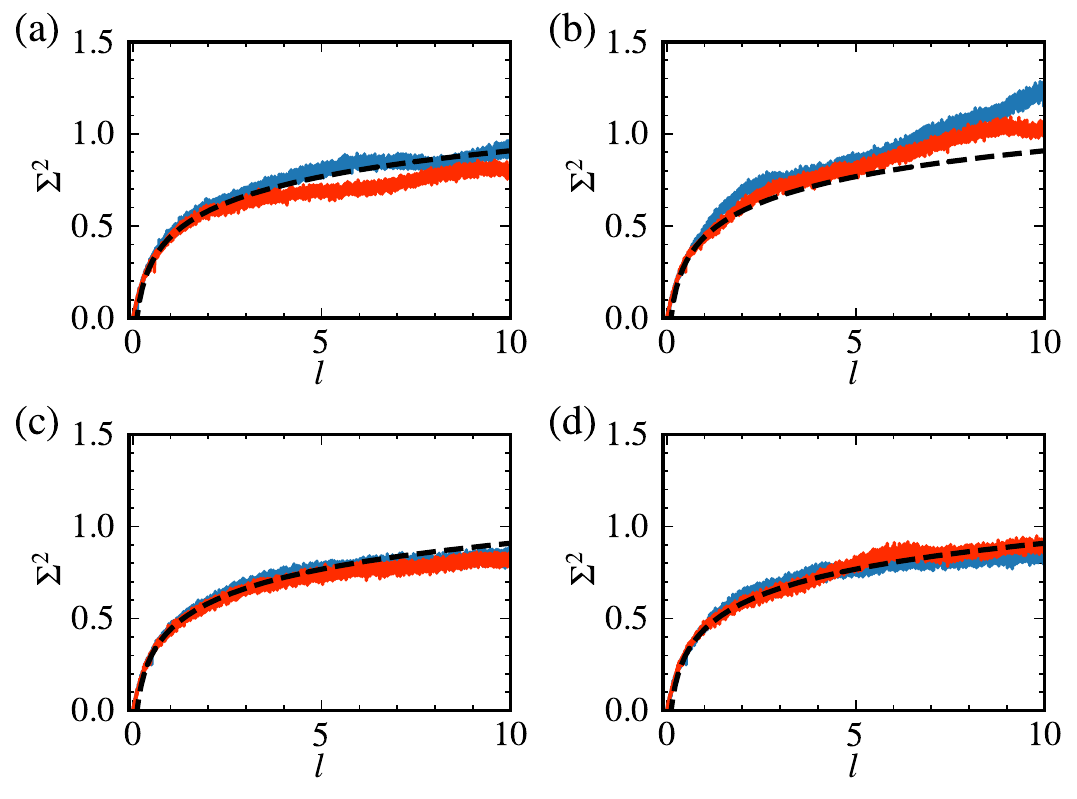}
		\caption{\label{Fig:LvlNoVar}Level number variance of singular values 
							of the	weights for 
							(a) the second hidden layer of MLP1024,
							(b) the second convolutional layer in the CNN miniAlexNet, 
							(c) the	second fully connected layer in AlexNet, and 
							(d) for the third dense layer in VGG19. Red curves show the 
							results for initialized weights and blue lines depict level 
							number variances for fully trained networks. The dashed, 
							black lines depict the theory prediction 
							Eq.~\eqref{Eq:SigPred} for the GOE.
							In all cases, the level number variance   
							grows logarithmically even after training. Particularly for 
							large networks in (c) and (d), where the statistics are most 
							reliable, deviations from the RMT prediction are small.}
	\end{figure}

\section{Level number variance } \label{Sec:LNV}
		The nearest neighbor  spacing distribution investigated above probes the 
		level statistics locally.  In order to probe long-range correlations between 
		singular values, we compute the level number variance Eq.~\eqref{Eq:LNV}, 
		which is considerably more sensitive to details of the singular value 
		distribution \cite{Guhr.1998,Mehta.2004}.
		To determine the level number variance numerically from the unfolded
		spectrum, for each $l$ we repeatedly draw values 
		$\xi_0\in[\min(\xi_i)+l/2,\max(\xi_i)-l/2]$ and count the number $n$ of $\xi_i$ in 
		the interval $[\xi_0-l/2,\xi_0+l/2]$. {These values are used to compute an estimate of the variance according to Eq.~\eqref{Eq:LNV}. }
		For random matrices from the Gaussian Orthogonal Ensemble 
		(GOE)	universality class, for large $l$ the level number variance is 
		given by 
		\cite{Guhr.1998,Mehta.2004,Weidenmuller.2009}
		\begin{align}
			\Sigma^2_{\rm GOE}(l) \approx \frac{2}{\pi^2}\left[\ln(2\pi l) 
			   +\gamma+1-\frac{\pi^2}{8}\right]\ ,
				\label{Eq:SigPred}
		\end{align}
		where $\gamma$ is the Euler-Mascheroni constant. It is known for GOE
		matrices that this formula is also a good approximation in the range of
		smaller $l$ \cite{Mehta.2004}, but one needs to keep in mind that  the  
		level number variance computed for empirical spectra depends on the  window 
		size $a$ chosen for broadening the spectrum. 
		We do not plot the level number variance for large $\ell\gg a$ as the unfolding procedure breaks down when the window size is much larger than the window over which the distribution is averaged. This limited plot range still allows to make the important distinction between a linearly growing level number variance (uncorrelated singular values) and logarithmic growth (singular values of a matrix with random bulk).

		We find good agreement of Eq.~\eqref{Eq:SigPred} 
		(dashed, black line in Fig.~\ref{Fig:LvlNoVar}) with the results for	
		trained networks (Fig.~\ref{Fig:LvlNoVar}) using a window size $a=15$. 
		In particular, there are only small differences between randomly 
		initialized weights (red lines in Fig.~\ref{Fig:LvlNoVar}) and the 
		fully trained weight matrices.

		\begin{figure}
			\centering
			\includegraphics[width=8.6cm]{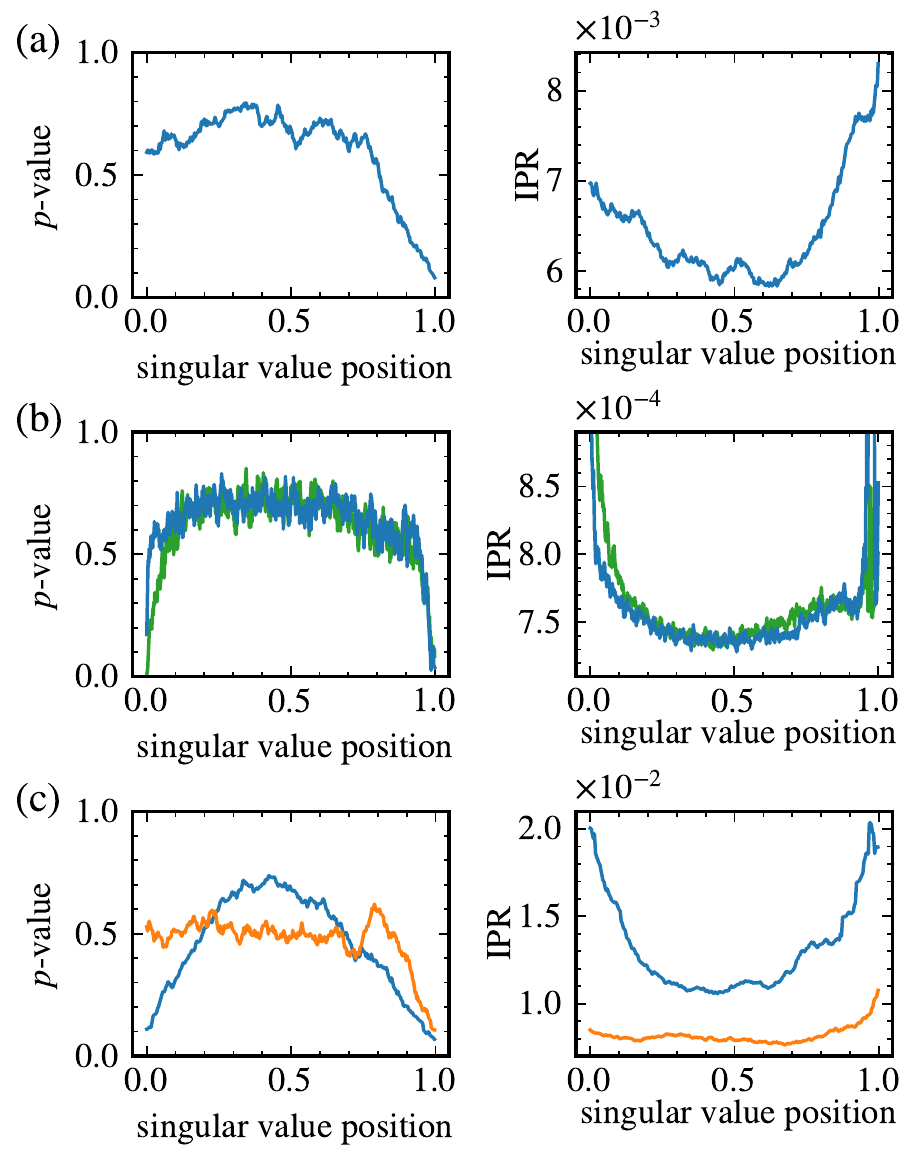}
			\caption{\label{Fig:EvecAnalysis}Analysis of the eigenvectors of 
							${\bf \sf W}^\dagger {\bf \sf W}$. 
							The left panels depict the $p$-values	of Kolmogorov-Smirnov 
							tests of the eigenvector entries versus a Gaussian 
							distribution. The right panels show the inverse participation 
							ratios of eigenvectors computed according to Eq.~\eqref{Eq:IPR}. 
							All results are averaged over neighboring eigenvectors with a 
							window size of 15. The $x$-direction describes the  position of 
							rank ordered singular values, such that 0 corresponds to the 
							smallest and 1 to the largest singular value of each weight 
							matrix. We show results for (a) the second hidden layer of
							{MLP1024}, 
							(b) the first dense layer of the large 
							pretrained DNNs 	
							AlexNet (blue) and VGG19 (green),
							and (c) the second convolutional layer (blue) {and first dense layer (orange)} of the CNN miniAlexNet. 
							The results for the $p$-values are consistent
							with those for the inverse participation ratio, indicating
							that relevant information is stored in eigenvectors corresponding
							to large singular values.}
		\end{figure}

\section{Distribution of entries of the singular vectors}

	  In addition to the singular values, we also analyze the singular 
		vectors of the weight matrices of trained DNNs. 
		For an $n\times m$ weight matrix ${\bf \sf W}$, we consider the 
		eigenvectors of ${\bf \sf W}^\dagger {\bf \sf W}$
		(right singular vectors) if $m\leq n$ or ${\bf \sf W} {\bf \sf W}^\dagger$ 
		(left singular vectors) otherwise.
		For a completely random	matrix of rank $m$, RMT predicts that the entries of 
		normalized eigenvectors follow a Gaussian distribution with zero mean and 
		standard deviation $\sigma=1/\sqrt{m}$.	We test for agreement between the 
		empirical distribution  of the $m$ entries of each singular vector with this RMT 
		prediction with the help of the Kolmogorov-Smirnov 
		test,   and average the resulting $p$-values over neighboring 
		rank ordered singular values with a window size of 15. 
		A large $p$-value indicates a random singular vector, as observed 
		for small and intermediate singular values for the MLP1024 
		network	and the large pretrained networks in the left panels of 
		Fig.~\ref{Fig:EvecAnalysis}(a) and (b), respectively. For large 
		singular values, the $p$-values decrease, suggesting that the 
		Gaussian hypothesis is rejected, and that information is stored in these
		vectors.  
		For the convolutional layer of the CNN in Fig.~\ref{Fig:EvecAnalysis}(c), we note that 
		eigenvectors corresponding to both small and large singular values show significant deviations from random behavior. 
		However, the 
		resulting matrix ${\bf \sf W} {\bf \sf W}^\dagger$ only has rank 300 
		which makes the analysis less reliable due to statistical	uncertainty.

		As a second measure for the randomness of singular vectors, we study their 
		degree of localization as measured by  the inverse participation ratio (IPR) 
		\cite{Guhr.1998,papenbrock2007colloquium,Weidenmuller.2009}
		\begin{align}
			{\rm IPR}(\bm{v}) = \sum_{i=1}^m |v_i|^4 \label{Eq:IPR}  \  \ .
		\end{align}
		In order to get some intuition for the IPR, we consider two examples: i) for 
		a normalized uniform $m$-dimensional vector with equal entries $1/\sqrt{m}$, 
		the IPR is given by $1/m$, the inverse of the number of relevant components. 
		In the case ii) of a vector with only one non-zero entry, the IPR is equal 
		to unity, again the inverse of the number of relevant components. 
		Since eigenvectors of GOE matrices have many relevant components, the IPR  
		random vectors is on the order of $1/m$, while a larger value indicates 
		deviations from RMT and the presence of	 learned information. This analysis 
		is in very good agreement with that of the $p$-values (see right panels in 
		Fig.~\ref{Fig:EvecAnalysis}) for large singular values. 	

	\begin{figure}[t!]
	\centering
	\includegraphics[width=8.6cm]{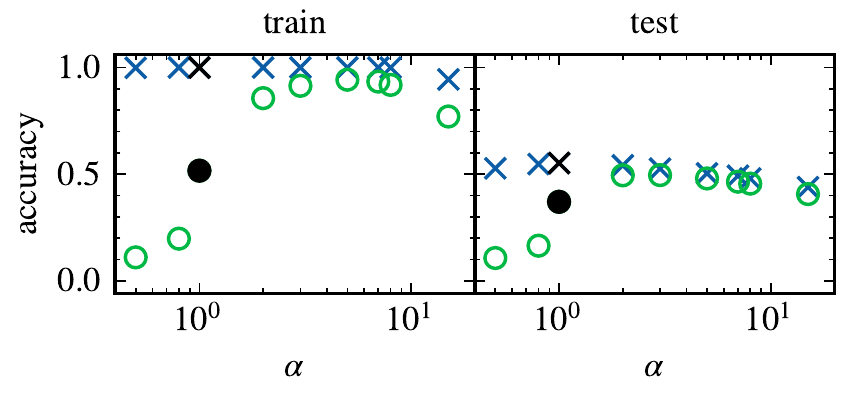}
	\caption{\label{Fig:Laziness}%
	    {Comparison of training and test accuracies for full MLP512 networks (blue crosses) and linearized networks (green circles) around the initial weights of the second layer as a function of the laziness hyperparameter $\alpha$. 
     The black symbols indicate accuracies for $\alpha=1$, similar to the training presented in Sec.~\ref{Sec:SummaryOfResults}.
     For small $\alpha<1$ accuracies of linearized and full networks deviate significantly which indicates rich learning, while for large $\alpha>1$ performance differences are small indicating lazy learning. 
		}}
    \end{figure}

	\begin{figure*}[t!]
	\centering
	\includegraphics[width=17.0cm]{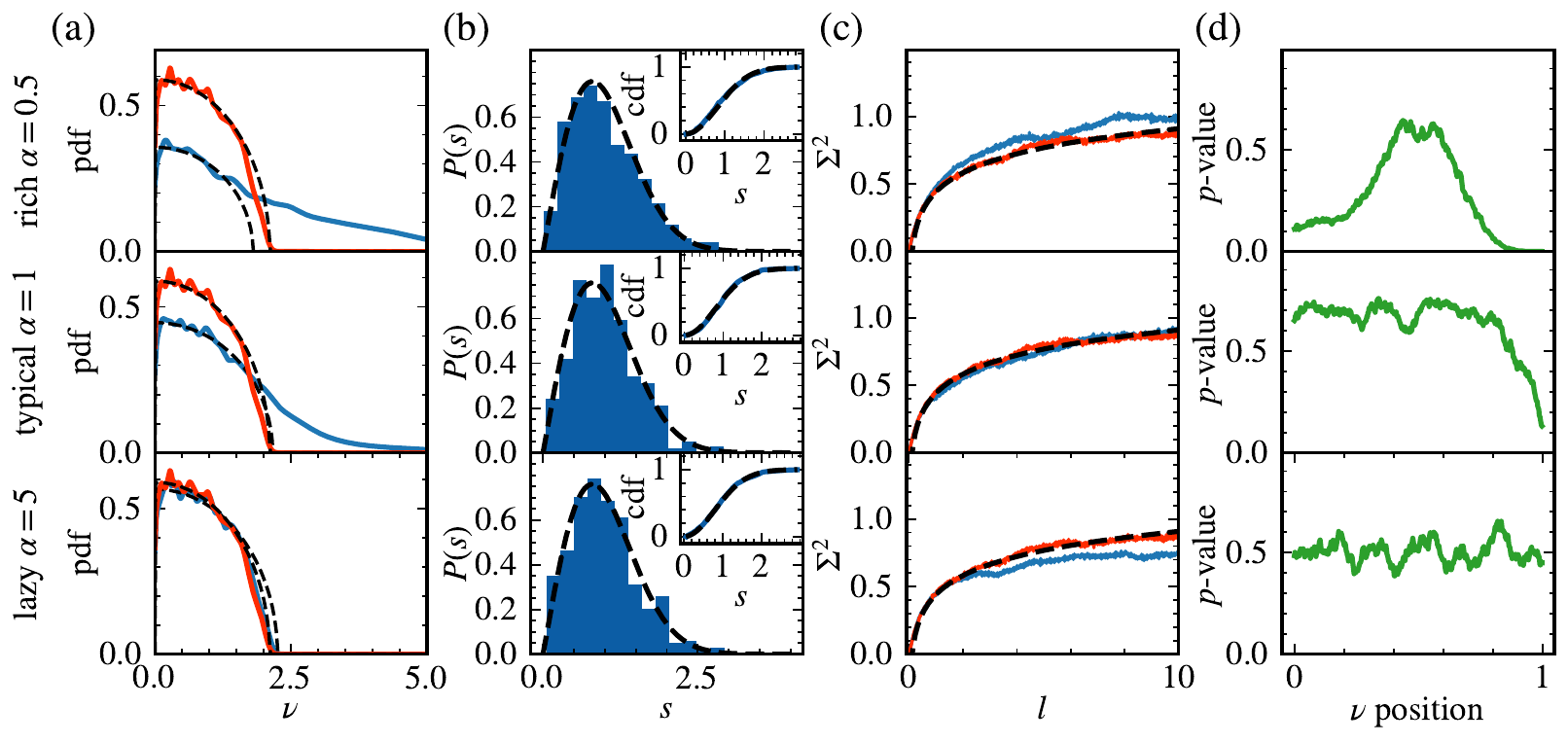}
	\caption{\label{Fig:LearningRegimes}%
	    {Random matrix theory analysis of second layer weights of MLP512 networks trained in different learning regimes: rich learning ($\alpha=0.5$, top panel), typical learning ($\alpha=1$, middle panel), and lazy learning ($\alpha=5$, bottom panel). We show (a) the spectra for trained (blue) and randomly initialized networks (red) together with fits of modified Marchenko-Pastur laws (dashed, black), (b) unfolded level spacing distributions (main panel, blue, window size 15) and corresponding cumulative distributions (insets) together with the Wigner surmise (dashed, black), (c) unfolded level number variance (trained: blue, initialized: red), and (d) p-values for a comparison of singular vector entries to a Porter-Thomas distribution. 
	    While trained networks in all cases follow universal RMT predictions (b and c), indicating a random bulk, 
	    lazy networks can be distinguished from typical and rich networks by the spectral distributions in (a) and p-values in (d).
		}}
    \end{figure*}

\section{RMT analysis of different learning regimes} \label{Sec:Lazy}
    {It was shown \cite{Du.2018,Li.2018,Du.2019,Allen.2019,Zou.2020,Zhang.2021} that neural networks can  achieve good generalization accuracies even when their weights change only by very small amounts during training. The opposite to this \emph{lazy learning} is denoted as \emph{rich learning}, where the final weights $\bm{W}$ after training deviate significantly from the initial ones $\bm{W}_0$. Based on this, a criterion for estimating the learning regime was proposed by Chizat et al.~\cite{Chizat.2019}: For a neural network $f_{\bm{W}}$ 
    that maps an  input $\bm{x}$ to an output, and an accuracy function $\mathcal{A}(f_{\bm{W}},\{\bm{x}\},\{\bm{y}\})$, where $\{\bm{x}\}$  is a dataset with labels $\{\bm{y}\}$, one computes the network's linearization around the initial weights $\bm{W}_0$
    \begin{align}
        \tilde{f}_{\bm{W}}(\bm{x}) &= {f}_{\bm{W}_0}(\bm{x}) + (\bm{W} - \bm{W}_0)\cdot \nabla_{\bm{W}} f_{\bm{W}} |_{\bm{W}_0} (\bm{x}) \ .
    \end{align}
    In the lazy learning regime, where $\bm{W}\approx \bm{W}_0$, linearization is a good approximation such that the accuracies are barely different, i.e.~
    \begin{align}
       \mathcal{A}(f_{\bm{W}},\{\bm{x}\},\{\bm{y}\}) &\approx \mathcal{A}(\tilde{f}_{\bm{W}},\{\bm{x}\},\{\bm{y}\}) \ .
    \end{align} 
    On the contrary, in the rich learning regime, one expects significant deviations such that 
    \begin{align}
       \mathcal{A}(f_{\bm{W}},\{\bm{x}\},\{\bm{y}\}) &\gg \mathcal{A}(\tilde{f}_{\bm{W}},\{\bm{x}\},\{\bm{y}\}) \ .
    \end{align}
    This criterion has the advantage that it can also be studied on a layer-wise basis by linearizing around a single weight matrix only, and as accuracies are in the range $[0,1]$, it gives a scale for laziness comparable between different network architectures. A disadvantage is that it requires to compute the linearization which can be resource-intensive for large networks. For obtaining $\tilde{f}_{\bm{W}}$, we use the \texttt{autodiff} implementation in the \texttt{jax} python package together with the \texttt{neural\_tangents} package \cite{neuraltangents2020}.
    }

    {We train several MLP512 networks, where laziness is controlled by introducing a hyperparameter $\alpha$ that modifies the output activations via \cite{Chizat.2019} 
    \begin{align}
     a_L &= \mathrm{softmax}\left(\alpha ({\bf \sf W}_{L}\bm{a}_{L-1} + \bm{b}_L)\right)   
    \end{align}
    and the cost function as
    \begin{align}
		l(\bm{W},\bm{b}) = -\frac{1}{N \alpha^2}\sum_{k=1}^N  \bm{y}^{(k)}
		\cdot\ln(\bm{a}_{\rm out}^{(k)}) \ .
	\end{align}
	Here, a large $\alpha>1$ scales down the gradient updates and therefore encourages lazy learning, while small $\alpha<1$ steers training towards the rich learning regime \cite{Chizat.2019}. 
	In the following, we focus on the second hidden layer of MLP512 networks considered in Sec.~\ref{Sec:SummaryOfResults} (see App.~C for other layers).
    Fig.~\ref{Fig:Laziness} depicts training accuracy (left panel) and test accuracy (right panel) as a function of the laziness parameter $\alpha$. As expected, the networks are in the rich regime for $\alpha<1$, where the full networks (blue crosses) perform significantly better than the linearized networks (green circles), while we observe lazy learning for $\alpha>1$. 
    The network with $\alpha = 1$ (black symbols), lies about in the middle between the two regimes, where we also find the best test accuracy.
    We therefore denote $\alpha=1$ as \emph{typical learning}.
    }
    
    {A comparison between the RMT analysis in the three regimes, rich $\alpha=0.5$ (top panel), typical $\alpha=1$ (middle panel), and lazy $\alpha=5$ (bottom panel) in Fig.~\ref{Fig:LearningRegimes} reveals: 
    \begin{enumerate}[label=(\roman*)]
    \item For all networks, the bulks of the spectra are random such that the universal properties, i.e.~level spacings (panel b) and level number variance (panel c) agree with RMT predictions. {By comparing the level number variance curves for networks trained with various $\alpha$ (not shown), we confirm that the level number variances in panel c display typical deviations from the RMT prediction such that there is 
    no trend of slower grow for networks with larger\;$\alpha$. }
    \item The rich network has more large singular values compared to the typical one, while the lazy network has almost an unchanged Marchenko-Pastur spectrum (panel a). It is therefore surprising that it still achieves a respectable test accuracy of $50.4\%$, compared to $52.7\%$ for the rich network and $55.2\%$ for the typical network. 
    \item The $p$-values for Kolmogorov-Smirnov tests of the eigenvector entries against a Porter-Thomas distribution (panel d) are small only for large singular values in the typical case. In the rich case, we observe small $p$-values also for singular vectors corresponding to the smallest singular values, and for the lazy network all $p$-values fluctuate only slightly around $0.5$ as expected for random weights.
    \end{enumerate}
    }
    
    These findings indicate that networks trained in the lazy regime do not deviate from RMT predictions after training, in striking contrast to 
    rich and typical networks. 
    Thus, an analysis of   singular value spectra and  singular vector entry distributions can be used to estimate the learning regime of a network on the level of the individual weights, without the need for a potentially resource intensive linearization of the networks.
    
    The almost perfect agreement with RMT predictions in the lazy regime raises the question whether the information that allows the network to still  generalize quite well is  encoded in parts of the spectrum that follow bulk statistics. If this was the case, it would seem  impossible to locate this information by means of an RMT analysis. However, we argue in the following that this is in fact not the case for the networks considered here. However, the RMT analysis in the lazy regime faces two major obstacles that make it difficult to detect the information hidden by 
    the dominant random bulk: 
    (i) an individual layer in the lazy network might carry very little information, as it is for example the case for the second hidden layer of the lazy MLP512 network shown in Fig.~\ref{Fig:LearningRegimes}, where by replacing the final weight with the initial one for this layer, the generalization accuracy of the network only drops from 50.4\% to 42.3\%. It is therefore not surprising that no extended region in the spectrum containing information is found by the RMT analysis. (ii) In the lazy case, the difference $\delta W=W-W_0$ between initial weight $W_0$ and the final weight after training $W$ is small, i.e.\@ ${\left||\delta W|\right|}/{||W_0||}\ll 1$. In this case, the sensitivity of RMT is not sufficient to detect signatures of $\delta W$ in $W$: For instance, the crossover from the Gaussian unitary ensemble to Poisson statistics has been studied in Ref.~\cite{Guhr.1997}, and it turns out that $\delta W$ would need to be much larger than in our case in order to have a noticeable effect compared to the bulk statistics from $W_0$. 
    
    As a solution, we suggest analyzing the 
    statistical properties of the  difference matrix $\delta W$ instead of the full weight  matrix. Such an analysis indeed indicates that $\delta W$ consists of a random bulk and a low-rank contribution that encodes relevant information (see App.~C).

	\begin{figure}[t!]
		\centering
		\includegraphics[width=8.6cm]{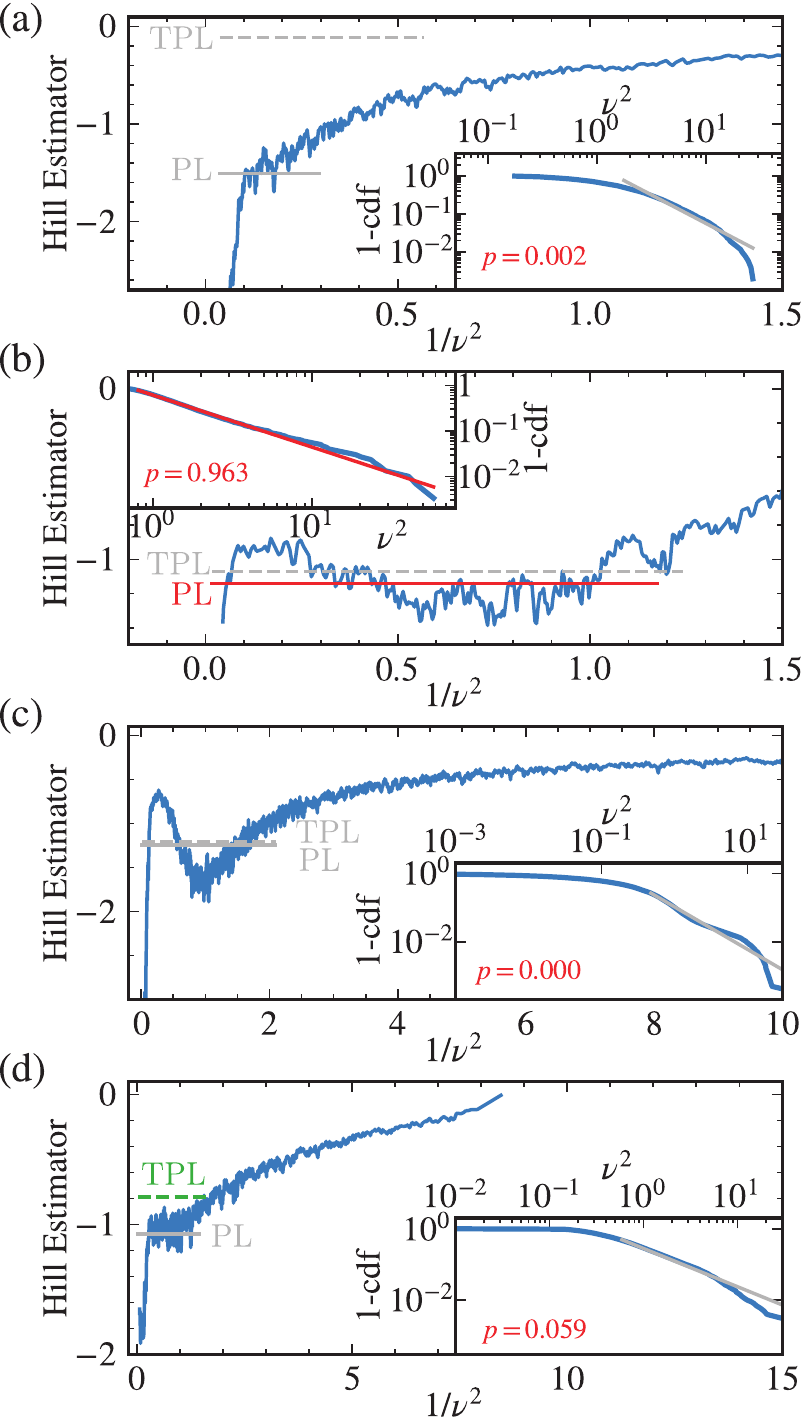}
		\caption{\label{Fig:Hill_estimator}%
						Analysis of the cumulative distribution of DNN singular values in the tail region. 
						Hill estimators of the weight matrix spectra for (a) the second 
						layer weight matrix of {MLP1024}, (b) the second 
						convolutional layer of the CNN miniAlexNet, (c) the second fully connected layer of AlexNet, and (d) the third fully connected layer of VGG19. The Hill estimators are obtained using 
						Eq.~\eqref{Eq:meanEvals} and Eq.~\eqref{Eq:meanHill} with a window
						size $a=20$. The insets depict the corresponding log-log cumulative
						distributions (blue). 
						{In addition, we show results of a power law fit $p(x)\propto x^{-\alpha}$ (solid gray or red) and a truncated power law fit $p(x)\propto x^{-\alpha} e^{-\lambda x}$ (dashed gray or green) to the tails, 
						where a power law tail corresponds to a flat  Hill estimator at $-(\alpha-1)$  starting 
						from $\nu_{\rm min}$.} {We show rejected fits in gray and accepted fits colored, based on p-values \cite{Clauset.2009} for the power laws and log-likelihood ratio tests versus truncated power laws (see also Table III in the appendix).}
																}
	\end{figure}    

\section{Hill estimator for tail spectra} \label{Sec:Hill} 	
		In previous work  \cite{Martin2021b,Martin.2021}, it has been argued that the spectral density of the singular values of  DNN weight matrices can be fitted by a power law. In order to study the distribution of large singular values further,   we employ the Hill estimator for power law tail exponents \cite{hill1975}, which  has been widely used in the applied finance, economics, and statistics literature \cite{akgiray1988,cheng1995,gopikrishnan.1999,lux1996,quintos2001,chan2007,hill2010}. In addition, we also study $p$-values of power law fits and find agreement with the results based on Hill estimators.

 For this purpose, we first rank order the eigenvalues $\lambda_i=\nu_i^2$ of the 
		weight matrices and compute the corresponding cumulative distribution function as
		\begin{align}
		    F(\lambda_i)= \frac{i}{N}
		\end{align}
		where a small index corresponds to a small eigenvalue.
		The Hill estimator $h$ is then 
		obtained from the local inverse slopes 
		\begin{align}
			\zeta_i = \frac{-\ln[\lambda_{i+1}/\lambda_i]}{\ln[F(\lambda_{i+1})/F(\lambda_i)]}
		\end{align}
		by averaging over $a$ surrounding eigenvalues 
		\begin{align}
			\tilde{\lambda}_i &= \frac{1}{a}\sum_{j=-a/2}^{a/2} \lambda_{i+j}\label{Eq:meanEvals} \\
			h(\tilde{\lambda}_i) &= \left(\frac{1}{a}\sum_{j=-a/2}^{a/2} \zeta_{i+j}\right)^{-1}\label{Eq:meanHill}
			\ .
		\end{align}
		The Hill estimator is sensitive to changes in the slopes of the log-log 
		cumulative distribution. In the presence of a power law tail, the value of the Hill estimator depends only weakly on the eigenvalue range over which it is determined, and the extrapolation $1/\lambda^2 \to 0$ yields the tail exponent of the distribution.  According to this criterion,  we do not find evidence of power law 
		tails for either i) MLP1024,  ii) miniAlexNet, nor iii) 
		AlexNet (Fig.~\ref{Fig:Hill_estimator} a-c). Only in the case of VGG19 
		(Fig.~\ref{Fig:Hill_estimator}d) there is a  region of intermediate singular values in the Hill plot with 
		a power law exponent of approximately one, but in the asymptotic regime of large singular values the exponent drops to approximately two, invalidating the concept that a single power law characterizes the distribution. 
		
		{To substantiate these findings, we additionally
		fit power laws  $p(x)\propto x^{-\alpha}$ to the spectra, compute their respective $p$-values using the algorithm by Clauset et al. \cite{Clauset.2009}, and compare the hypothesis of a power law to a truncated power law $p(x)\propto x^{-\alpha} e^{-\lambda x}$ with log-likelihood ratio tests (see also App.~D). 
		The fitting procedure and the comparison of the power law hypothesis to other distributions is implemented in the \texttt{powerlaw} package while the $p$-values are computed by numerical approximation of the test statistic with $2500$ synthetic power law fits followed by a Kolmogorov-Smirnov test as described in \cite{Clauset.2009}.}
    Out of the four spectra analyzed, the Hill plot does not look consistent with a straight line in cases a), c), and d), while case b) fluctuates around a straight line. Indeed,  for  AlexNet (panel c) and MLP512 (panel a) the $p$-values reject the power law hypothesis. For the third dense layer in VGG19 (panel d), a power law cannot be rejected based on $p = 0.059$, slightly larger than the threshold. From the likelihood ratio test 
   however,  we find better agreement for a truncated power law (in agreement with \cite{Martin.2021})

	In summary, three out of the four examples analyzed here are not described by power law tails. In Appendix D Table III we present results for additional networks, and  in only two out of eight cases a power law cannot be rejected. Hence, we conclude that there is no evidence that the singular values of DNN weight matrices are generically described by a power law tail distributions, and argue that the exponent resulting from a power law fit to the singular value probability density function can only be viewed as a heuristic tool to characterize different spectra but not a genuine property of the tail of the distribution of singular values. In addition, given the absence of power law tails in {most of} the singular value distributions {studied here}, it is unclear whether  weight matrices of fully trained DNNs indeed belong to an ensemble of heavy tailed random matrices as suggested in \cite{Martin2021b}.

\section{Conclusions}
	
	The complexity of over-parameterized DNNs   
	makes it difficult to understand
	their learning and generalization behavior.
	In light of this, we have applied RMT as a zero information hypothesis in
	order to separate randomness from information. 
	In particular, since at initialization weights are chosen randomly from a
	probability distribution, the corresponding weight matrices agree 
	perfectly with predictions of RMT before training. 
	Specifically, the singular value spectra of initialized networks are 
	governed by a Marchenko-Pastur distribution, the level spacing 
	distribution follows the Wigner surmise, and the level number variance 
	only grows logarithmically.  By comparing these characteristics between 
	randomly initialized and trained networks, one can understand which parts 
	of the weight matrix singular value spectrum stores information. 
	This approach works  well for image classification problems where we find that the underlying rule in the dataset is stored as a low-rank contribution in the trained weights. It remains an open question to what extent this can be applied to networks trained for other types of problems.
	  
	We find that even in fully trained DNNs large parts of the eigenvalue spectrum remain random. 
	In particular,  we demonstrate that  the agreement between the level 
	spacing distribution of the bulk of singular values for fully trained 
	networks and the parameter free Wigner surmise is excellent, and
	that the even more sensitive level number variance continues to agree 
	with the RMT prediction as well.
	In agreement with the spectra, an analysis of the singular vectors 
	reveals that they are also predominantly random, except for the ones 
	corresponding to the largest singular values.
	This shows that the majority of the weight matrix does not contain 
	relevant information, and that learned information may be concentrated in the largest 
	singular values and associated vectors only.
	Interestingly, we find the strongest deviations from RMT predictions for networks trained in the so-called rich regime, while for networks trained in the lazy regime the whole spectrum and all eigenvectors follow RMT results. Since the best 
	generalization performance for learning is observed in between lazy and rich regime, the ability to  efficiently determining the learning regime on a layer-wise basis using the singular value spectra and singular vectors could allow for algorithms that dynamically steers the training towards the desired regime.
		An  analysis of the tails of the eigenvalue spectra using the Hill estimator shows that the distributions are heavy tailed but there is 
	in general no evidence for a single power law in the tail region of the distribution 
	in general.

\begin{acknowledgments}  
	 This work has been funded by the Deutsche Forschungsgemeinschaft (DFG) 
	 under Grants No.~RO 2247/11-1 and No.~406116891 within the Research 
	Training Group RTG 2522/1.\\ 
\end{acknowledgments}

	 M.~T.~and M.~S.~contributed equally to this work.\\

\renewcommand{\theequation}{A\arabic{equation}}
\setcounter{equation}{0} 
\section*{APPENDIX A: Details on neural networks} 
	\begin{table*}[t]
			\caption{\label{Tab:Networks}Neural network architectures and 
			performance of trained networks. We use d to indicate a dense
			layer, c for a convolutional layer, p for max pooling, f for
			flattening, and r for response normalization
			layer (with a depth radius of 5, a bias of 1, $\alpha=1$, and 
			$\beta=0.5$).}
			\begin{tabular}{|l|c|c|r|c|c|}
				\hline
				& network & dataset & 
				training acc & test acc \\
				\hline
				\multirow{1}{0.47cm}{i)} 	&
						\multirow{1}{12.5cm}{MLP512, seed 1 (Sec.~II)
							\{d 3072, d 512, d 512, d 512, d 10\} \cite{Zhang.2021}} & CIFAR-10
											& 
											100\% & 54.7\%\\  
				\hline		
				\multirow{1}{0.47cm}{} 	&
						\multirow{1}{12.5cm}{MLP512, seed 2
							\{d 3072, d 512, d 512, d 512, d 10\}  } & CIFAR-10
											& 
											100\% & 55.1\%\\  
				\hline	
				\multirow{1}{0.47cm}{} 	&
						\multirow{1}{12.5cm}{MLP512, seed 3
							\{d 3072, d 512, d 512, d 512, d 10\}  } & CIFAR-10
											& 
											100\% & 55.2\%\\  
				\hline	
				\multirow{1}{0.47cm}{ii)}	&
						\multirow{1}{12.5cm}{MLP1024
							\{d 3072, d 1024, d 512, d 512, d 10\} }  & \multirow{1}{1.45cm}{CIFAR-10}
											& 
											100\% & 55.4\%\\  
				\hline
				\multirow{1}{0.47cm}{iii)}& 
						\multirow{1}{12.5cm}{miniAlexNet
						  \{c 300 $5\times5$, p $3\times3$, r, c 150 $5\times5$,
							p $3\times3$, r, f, d 384, d 192, d 10\}  \cite{Zhang.2021}} & \multirow{1}{1.45cm}{CIFAR-10}
											&  
											100\% & 78.5\%\\  
				\hline	 			
				\multirow{1}{0.47cm}{iv)}	& 
						\multirow{1}{12.5cm}{AlexNet \cite{Krizhevsky.2017} -- \texttt{torch}} & ImageNet
										  & 
												& 56.5\%\\
				\hline		
				\multirow{1}{0.47cm}{v)}	& 
						\multirow{1}{12.5cm}{VGG16 \cite{Simonyan.2014} -- \texttt{tensorflow}}& ImageNet
											& 
												& 67.6\%\\
				\hline
			    \multirow{1}{0.47cm}{vi)}	& 
						\multirow{1}{12.5cm}{VGG19 \cite{Simonyan.2014} -- \texttt{torch}}& ImageNet
											&  
												& 72.4\%\\
				\hline
			\end{tabular}
		\end{table*} 
		In this article, we consider a variety of different networks to show that
		our results are valid for a wide range of architectures. 
		Tab.~\ref{Tab:Networks} lists the network architectures, training 
		datasets, and accuracies achieved on each dataset. We downloaded the 
		large pretrained networks iv) AlexNet \cite{Krizhevsky.2017} via 
		{\texttt{pytorch}} \cite{pytorch.2019},  v) VGG16 \cite{Simonyan.2014} via \texttt{tensorflow} \cite{Tensorflow.2015}, and vi) VGG19 
		\cite{Simonyan.2014} via {\texttt{pytorch}} \cite{pytorch.2019}.
		Networks i)-iii) are trained using mini-batch stochastic gradient decent
		for 100	epochs. The weights 
		are initialized	using the Glorot uniform distribution \cite{Glorot.2010}
		and the biases are initialized with zeros. We standardize each image of 
		the CIFAR-10 dataset by subtracting the mean and dividing by the 
		standard deviation. We set the learning rate to 0.001 at the beginning 
		and use an exponential learning rate schedule with decay constant 
		$0.95$.  
		For all networks, we choose 0.95 as momentum and the mini-batch size is
		32. 
		Network architectures i)-ii) in Tab.~\ref{Tab:Networks} are trained without $L_2$
		regularization, while we use an $L_2$ regularization strength of 
		$10^{-4}$	for training miniAlexNet networks iii). 

\renewcommand{\theequation}{B\arabic{equation}}
\setcounter{equation}{0} 
\section*{APPENDIX B: Reshaping of convolutional layer weights}
\begin{figure}[t!]
	\centering
	\includegraphics[width=8.6cm]{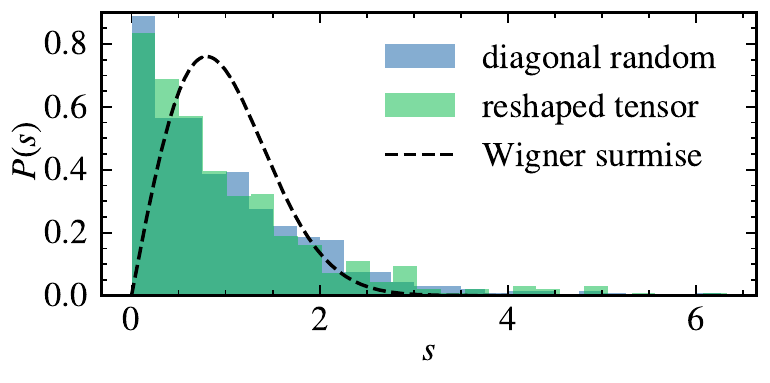}
	\caption{\label{Fig:CNNReshaping}%
					{Level spacing distribution of a diagonal i.i.d.\@ Gaussian 
					random matrix (blue) and for comparison for a tensor $10\times10\times600\times 600$
					where the  $100$ constituent  matrices are diagonal with  i.i.d.\@ Gaussian 
					random entries (green). To compute the singular value decomposition of the tensor, it is reshaped to size $60000\times 600$ as described in Eq.~\eqref{Eq:CNNreshape}. We find that reshaping conserves the Poisson statistics that is very different from the Wigner surmise (black).    } 	
		} 
\end{figure}

		In the case of convolutional layers we have to reshape the filters 
		before computing the singular value decomposition \cite{Yoshida.2017}. 
		{We unfold the four 
		dimensional	weights in such a way that the number of rows corresponds to the largest dimension of the weights.
		Each 
		row is then filled with the remaining three dimensions that are listed with the 
		smallest dimensions last to keep points that are close in the 3D 
		tensor also close in 1D.} In formal terms, this means that for a 
		convolutional layer $\sf W$ of the form $(K,L,M,N)$, where without loss of 
		generality $K\geq L\geq M\geq N$, the reshaped matrix $\tilde{\sf W}$ with dimensions 
		$(K,L\cdot M\cdot N)$ is related to the original matrix  in the following way:
		\begin{align}
			\tilde{\sf W}_{k, ( l\cdot M\cdot N+m \cdot N+n)} = {\sf W}_{k,l,m,n}   \label{Eq:CNNreshape}
		\end{align}
		All indices start at zero. While this procedure is to some degree 
		arbitrary we convinced ourselves that similar methods give the same
		qualitative results. We further checked that this reshaping preserves the
		structure of the original filters: we construct a 4D tensor of shape $10 \times 10 \times 600 \times600$ by drawing 100 times a diagonal, i.i.d., Gaussian $600 \times 600$ matrix, for which the level spacing distribution is a Poisson distribution. After reshaping, we find that the resulting level spacing distribution of the 2D matrix remains Poisson distributed (Fig.~\ref{Fig:CNNReshaping}). This is the theoretically expected distribution of a diagonal matrix \cite{Brody.1981}.

\begin{figure}[t!]
	\centering
	\includegraphics[width=8.6cm]{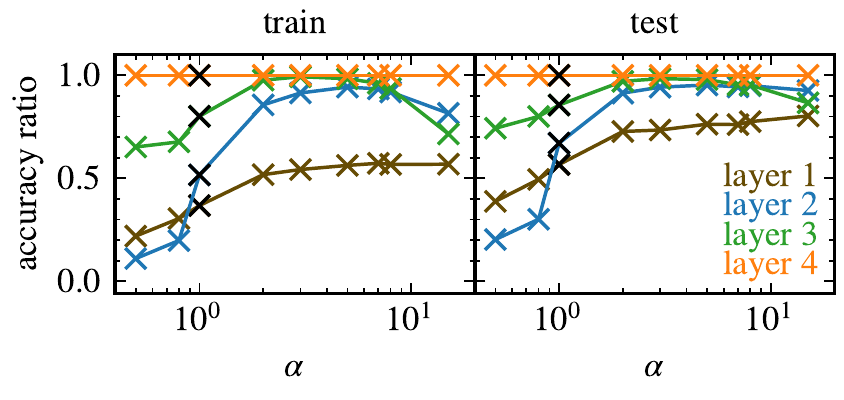}
	\caption{\label{Fig:LazinessOtherLayers}%
					{Comparison of training and test accuracies for full MLP512 networks and linearized networks around the initial weights of several layers (first layer: brown, second layer: blue, third layer: green, output layer: orange) as a function of the laziness hyperparameter $\alpha$. 
					Crosses show the ratio between linearized and full network accuracy and lines are a guide to the eye.
					The black crosses indicate accuracy ratios for $\alpha=1$, similar to the training presented in Sec.~\ref{Sec:SummaryOfResults}.     } 	
		} 
\end{figure}

\renewcommand{\theequation}{C\arabic{equation}}
\setcounter{equation}{0} 
\section*{APPENDIX C: Learning regimes of other MLP512 layers} \label{App:LeraningRegimes}	
    {We additionally compare test and training accuracies similar to main text Fig.~\ref{Fig:Laziness} for the other layers of the MLP512 networks trained with various laziness parameters $\alpha$. In Fig.~\ref{Fig:LazinessOtherLayers}, we show the ratio of the training accuracy between linearized and full networks, where a large ratio $\approx 1$ indicates lazy learning and small values indicate rich learning. We find that for the second layer, $\alpha$ interpolates nicely between lazy and rich regime, the first layer tends towards rich learning, and the weights for the third layer are closer to lazy learning. The weight of the output layer, which has only rank ten, is always lazy. We note that this imbalance between the layers could be lifted by using different learning rates per layer.}
 
 \begin{figure}[t!]
	\centering
	\includegraphics[width=8.6cm]{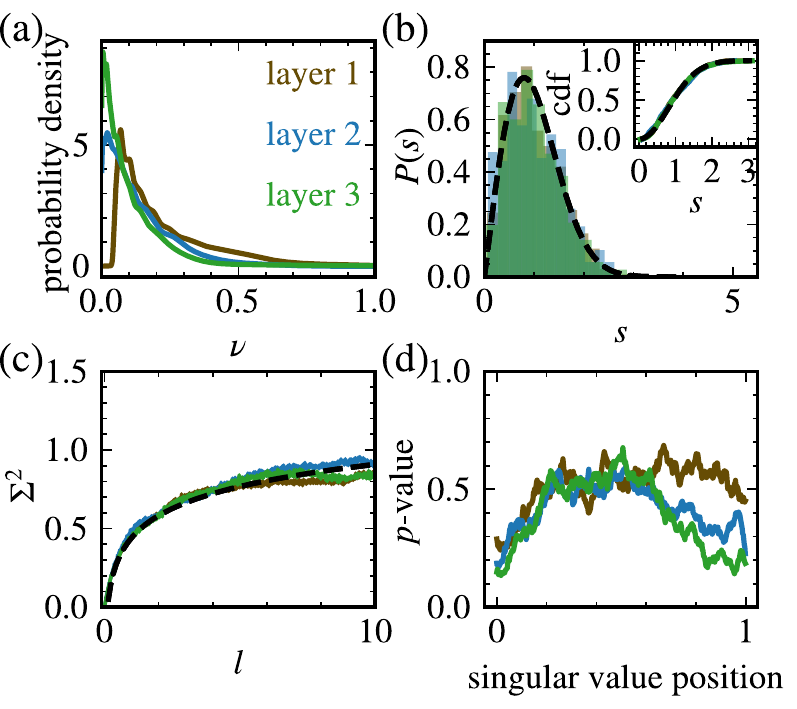}
	\caption{\label{Fig:RMTdWLazy}%
					{Random matrix theory analysis for the difference $\delta W^{(l)}=W^{(l)}-W^{(l)}_0$ between the trained weights $W^{(l)}$ and the initial weights $W^{(l)}_0$ for the lazy MLP512 network trained with $\alpha=5$. We show results for the first hidden layer (brown), second hidden layer (blue), and third hidden layer (green). (a) Singular value spectra obtained via Gaussian broadening. (b) Level spacing statistics of the unfolded singular value spectra. (c) Level number variance of the unfolded spectra. (d) Averaged $p$-values for a comparison between singular vector entries and the Porter-Thomas distribution with window size $2a=30$. The spectrum of $\delta W^{(l)}$
					agrees well  with the Wigner surmise and a logarithmicly growing level number variance. However,  the $p$-values in panel (d) indicate that again large parts of the spectrum of $\delta W^{(l)}$ are random and that information is stored in the largest singular values and corresponding vectors of $\delta W^{(l)}$. In addition,  the singular value spectrum is not of Marchenko-Pastur type.}  	
		} 
\end{figure}

    In the lazy learning regime the trained weights $W^{(l)}$ remain close to the initial random matrices $W^{(l)}_0$ such that the random bulk dominates any RMT analysis of $W^{(l)}$ and therefore masks small local deviations from the bulk statistics. We thus  consider  the deviations from the initial weights $\delta W^{(l)} = W^{(l)} -W^{(l)}_0$,  which  again have random bulk statistics.  Their unfolded spectra follow the Wigner surmise (Fig.~\ref{Fig:RMTdWLazy}a), and one finds a logarithmic increase of the level number variance (panel c). In contrast to the full weight matrix, the difference matrix has a qualitatively different distribution function of the singular values (panel a) and the $p$-values (panel d) are consistent with information being stored in parts of the spectrum corresponding to large singular values similar to what we found in the main text for networks in the typical learning regime.

\renewcommand{\theequation}{D\arabic{equation}}
\setcounter{equation}{0}   
\section*{APPENDIX D: Details on the tail distributions of weight spectra} \label{App:Tails}	 
{In addition to the results of Fig.~\ref{Fig:Hill_estimator}, we fit power laws to the tails of the other dense layers of the \texttt{pytorch} model of AlexNet and VGG19, also considered in \cite{Martin.2021}. We also provide the $p$-values for the power law fits \cite{Clauset.2009} and the results of log-likelihood ratio tests between truncated  power law $p_{\rm TPL}(x)\propto x^{-\alpha}e^{-\lambda x}$ and power law $p_{\rm PL}(x)\propto x^{-\alpha}$. The log-likelihood ratio for $n$ data points in the tail $x_i$ is defined as \cite{Clauset.2009}
\begin{align}
    R &= \frac{1}{\sqrt{2n}\sigma} \ln \prod_{i=1}^n \frac{p_{\rm TPL}(x_i)}{p_{\rm PL}(x_i)} \ ,
\end{align}
where $\sigma$ is the empirical standard deviation of the difference $\ln p_{\rm TPL}(x_i) - \ln p_{\rm PL}(x_i)$. A positive sign of $R$ thus indicates that a truncated power law is a better fit, and a negative sign indicates a better fit for a power law. The $p_2$ value is then defined as the probability to obtain a ratio with magnitude of at least $|R|$ from a distribution $p(R)$ centered at zero with standard deviation $\sigma$, i.e.~the probability that the sign is only due to fluctuations. Therefore, small $p_2<0.05$ indicate a reliable sign of $R$, while large $p_2$ indicate an unreliable sign from fluctuations and hence an inconclusive test.
For the dense layers of the large pretrained networks (see Tab.~III), we only accept a power law in a single case, the third dense layer of AlexNet. For the third dense layer of VGG19, a power law is not rejected, however, a truncated power law yields a better fit. For the other layers, already the power law is rejected based on $p<0.05$.}
 
\begin{table}[t]
			\caption{\label{Tab:PLfits}{Power law fit results and log-likelihood ratio tests between truncated power law and power law for the dense layers of AlexNet, VGG19, and the weights considered in Fig.~\ref{Fig:Hill_estimator}. Here, d denotes a dense layer and c a convolutional layer. We reject the power law if $p<0.05$ or in the case where the two-distribution test favors a truncated power law (positive $R$, $p_2<0.05$) \cite{Clauset.2009}.}}
			\begin{tabular}{|c|c|c|c|c|c|c|c|}
				\hline
				\multirow{2}{*}{network} & \multirow{2}{*}{layer} & PL fit	& 2-distr. & 2-distr. & PL fit & PL fit        & reject  \\
				                         &   & $p$     &  $R$       &  $p_2$       &$\alpha$& $x_{\rm min}$ & PL \\ 
			    \hline
			    MLP1024                 & d2      & 0.0019  & 2.50      & 0.000      & 2.51   &  3.389        & yes \\
			    \hline
			    {mAlexNet}             & c2      & 0.9628  & 1.02      & 0.363      & 2.14   & 0.852         & no \\
			    \hline
			    \multirow{3}{*}{AlexNet}& d1      & 0.0004  & 1.55      & 0.359      & 2.29   &  0.418        & yes\\ \cline{2-8}
			                            & d2      & 0.0004  & 1.25      & 0.211      & 2.25   &  0.480        & yes\\ \cline{2-8}
			                            & d3      & 0.9990  & -0.003      & 0.999      & 3.02   &  2.046        & no\\  
			    \hline
			    \multirow{3}{*}{VGG19}  & d1      & 0.0011  & 2.01      & 0.142      & 2.27   &  0.275        & yes\\ \cline{2-8}
			                            & d2      & 0.0007  & 1.98      & 0.055      & 2.19   &  0.291        & yes\\ \cline{2-8}
			                            & d3      & 0.0590  & 2.26      & 0.001      & 2.07   &  0.690        & yes\\  
			    \hline
	        \end{tabular}
\end{table}

\end{document}